\documentclass[12pt]{article}
\usepackage{graphicx}
\providecommand{\keywords}[1]{\textbf{\textit{keywords:\,\,}}#1}
\usepackage{amssymb}
\usepackage{cite}
\usepackage{mathtools}
\usepackage[a4paper,left=2cm,right=2cm,top=2cm,bottom=2.5cm] {geometry}
\begin{document}
%=================================================================
\title {\bf Dispersive reservoir influence on the
superconducting phase qubit}
%----------------------
%--------------------------------------------
\author{A. H. Homid$^{1,}\footnote{E-mail :ahomid86@gmail.com}$\,,
A.-B. A. Mohamed$^{2,3}$ and  H. A. Hessian$^{3}$
\\
\small$^{1}$ Faculty of Science, Al-Azhar University, Assiut,
Egypt\\
\small$^{2}$ AL-Aflaj Community College, Salman Bin Abdulaziz University, Saudi Arabia\\
\small$^{3}$ Faculty of Science, Assiut University, Assiut, Egypt}
\date{}
%wwwwwwwwwwwwwwwwwwwwwwwwwwwwwwwwwwwwwwwwwwwwwwwwwwwwwwwwwwwwwwwwwwwwwwwwwwwwwwwwwwwwwwwwwwwww
\maketitle
%------------------------
\begin{abstract}
%------------------------------------------
In this article, an analytical description is presented based on the
master equation.
%-----------------------------------------------
This master equation is formed from the system of superconducting
phase qubit which is coupled to a torsional resonator and damped by
a dispersive reservoir.
%---------------------------------------------
An analytical approach for searching of some physical phenomena on
the qubit system is presented, such as qubit inversion, purity and
negativity.
%------------------------------------------
In addition, these  phenomena are discussed for the resonance and
off-resonance cases for many different initial states.
%-------------------------------------
From computational results, it is found that the mentioned phenomena
depend on the dispersive reservoir parameter which leads to their
death.
%------------------------------------------------
However, the complete destruction of system coherence in the
presence of dispersive reservoir leads to the death of entanglement
between the qubit and torsional resonator.
%---------------------------------------------
\end{abstract}
%=========================================================================
%=======================================================================
\section{Introduction}
%------------------------
\hspace{0.5cm}
%-------------------------------
In the recent years, the quantum entanglement and mixedness are two
biggest challenges of quantum information theory.
%--------------------------------------------------------------------------
Quantum entanglement is a very important quantum characteristic that
has no classical counterpart.
%-------------------------------------------------------------------------
Also, quantum entanglement is necessary for the realization of
quantum communication and the most important computational tasks,
such as quantum teleportation \cite{b1}, superdense coding \cite{b2}
and so on.
%--------------------------------------------------------------------------
In addition, it is crucial to understand the fundamental aspects of
quantum mechanics, including quantum measurement and quantum
decoherence \cite{b3}.
%--------------------------------------------------------------------------
Consequently, no system can be completely isolated from its
environment.
%--------------------------------------------------------------------------
Thus in real experimental situations due to the coupling
environment, the entangled state inevitably loses its purity and
becomes mixed.
%-------------------------------
%========================================================================
\\ \indent
%--------------------------------
Decoherence phenomenon is the most dangerous obstacle for all
entanglement manipulations.
%------------------------------------------
However, the recent studies of decoherence effect for different
systems are increased due to the variety of applications in areas as
quantum optics and quantum information \cite{ad1,ad11,ad2,ad3}.
%----------------------------------------------------------
In fact, the interactions between a quantum system with the
environment are caused two types of irreversible effects, e.g.,
quantum dissipation and quantum decoherence \cite{b4,b5}.
%---------------------------------------------------
The former will cause energy dissipation, and the latter will make
the system degenerate from coherent state to classical state.
%--------------------------------------------------------------------------
Entanglement evolution under decoherence has attracted much interest
and many researchers have studied it extensively based on various
models in the view of environment-induced decoherence \cite{b5d,b5d1}.
%==================================================================================
\\ \indent
%-----------------------
In the previous years, some researchers have found that state
entanglement can be decreased to zero abruptly and remained zero for
a finite time.
%--------------------------------------------------------------------------
This phenomenon is called entanglement sudden death
\cite{b6,b7,b8,bb8,bbb8}, where entanglement of two-qubit decays to
zero in a finite time under the influence of pure vacuum noise
\cite{b9}.
%--------------------------------------------------------------------------
Precisely, entanglement sudden death has been experimentally
observed in an implementation using twin photons \cite{b10} and
atomic ensembles \cite{b11}.
%--------------------------------------------------------------------------
In many efforts have been devoted to study various types of
superconducting (SC) qubit which is considered as good candidate to
solve this problem.
%--------------------------------------------------------------------------
%==========================================================================
\\ \indent
%---------------------------
Nowadays, superconductors are viable elements and feasible within
current experimental technology.
%------------------------------------------------------
SC qubit is represented a unique and its interesting approach to
quantum information and quantum computation because it naturally
allow strong coupling.
%-----------------------------------------------------
Some aspects which are discussed through these systems: entanglement
between a flux qubit and a superconducting quantum interference
device (SQUID) \cite{b12} have been realized experimentally.
%---------------------------------------------------------------
In addition, there are other aspects: the strong coupling of a
single photon to a SC qubit has been experimentally demonstrated
\cite{b13} by using a one-dimensional transmission line resonator
\cite{b14}.
%---------------------------------------------------------------
Recently, some quantum logic gates have proposed \cite{b14d1}
by using a system consisting of a number of SC-charge qubits
coupled to a resonator to study different quantum operations.
%-----------------------------------------------------------------------
Also, a new physical scheme for implementing a discrete quantum
Fourier transform is proposed via SC qubits coupled to a single-mode
SC-cavity \cite{b14d2}.
%---------------------------------------------------------------
However, the effect of environment cannot be neglected in these
situations \cite{b15}, because, the qubits are never isolated, but
under some environmental influence.
%=================================================================
\\ \indent
%------------------------------
Our purpose of this article is to give an analysis of the
entanglement dynamics of a SC phase qubit coupled to a torsional
resonator.
%---------------------------------------------------------
Consequently, some phenomena, such as the qubit inversion, purity
and negativity are theoretically investigated, mainly because such
states are more physical and experimentally relevant.
%--------------------------------------------------------
Thus we take into account decoherence and analyze its effect on
these phenomena.
%--------------------------------------------------------
Despite the complexity of the problem, the analytical descriptions
of the exact solution of the master equation is presented.
%=================================================================
\\ \indent
%----------------------------------
This article is arranged as follows: The physical system of qubit
with torsional resonator and its dynamics  is provided in section 2.
%---------------------------
The influence of  phase-damping on a SC qubit inversion and measures
of entanglement is presented in section 3.
%--------------------------------------------------
Finally,the conclusions is demonstrated in section 4.
%wwwwwwwwwwwwwwwwwwwwwwwwwwwwwwwwwwwwwwwwwwwwwwwwwwwwwwwwwwwwwwwwwwww
\section{The physical system of qubit with
torsional resonator and its dynamics }
\begin{figure}
\centering\includegraphics[height=10cm,width=14cm]{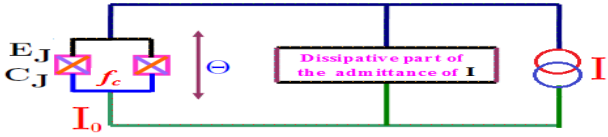}
\caption{Circuit diagram of the phase qubit; a single Josephson
junction with Josephson coupling energy $E_{J}$ and Josephson
capacitance $C_{J}$ is driven with a bias current $I$. The rectangle
in this diagram refer to dissipative  part of the admittance of the
bias current $I$.}\label{w1}
\end{figure}

%-----------------------------
\hspace{0.5cm}
%-------------------------------
We introduce a theoretical description of the superconducting phase
qubit coupled to a torsional resonator.
%==============================
A phase qubit with $\kappa$ excess Cooper-pair charges consists of
two superconducting islands connected to a superconducting electrode
through two identical Josephson tunnel junctions of a  small size
with the same capacitance $C_{J}$, see Fig.\ref{w1}.
%---------------------------------
If the phase qubit is working in a regime with $k_{B}T\ll E_{C}\ll
E_{J}$, where $E_{J}$, $E_{C}$, $k_{B}$ and $T$ are Josephson
energy, charging energy, Boltzmann constant and temperature,
respectively.
%------------------------------------------------------------
The Hamiltonian of the superconducting phase qubit can be described
by the form \cite{b16}
%------------------
\begin{eqnarray}\label{a}
\hat{H}_{qu}&=& \frac{(2e)^{2}}{2C_{J}}\kappa^{2}-\frac{\hbar
I_{0}}{2e}\Big(\cos(\pi f_{c})+\frac{
I}{I_{0}}\Theta\Big)\cos(\Theta-\pi f_{c}),
\end{eqnarray}
%-----------------------------------------------------
where $\Theta$, $I$, $I_{0}$, and $f_{c}$ are the phase difference
across the junction, bias current, junction critical current, and
the external flux which is generated by a classical applied magnetic
field, respectively.
%-----------------------------------------------------------------------
Moreover, $\kappa$ and $\Theta$ are quantum mechanical conjugate
variables and satisfy $[\kappa,\Theta]=-i$.
%--------------------------------------------------------
\\ \indent
%-------------------------------
In this study, the bias current $I$ is assumed to be zero.
%----------------------------------
The Josephson energy, and the charging
energy for each junction also are assumed to be $E_{J}=\hbar
I_{0}/2e\sim6.8\times10^{-21}\,\,J$  and
$E_{C}=\frac{(2e)^{2}}{2C_{J}}\sim8.3\times10^{-27}\,\,J$  as estimated in \cite{c1}.
%-----------------------------------------------------
Therefore, the effective Josephson coupling $E_{J_{e}}=E_{J}\cos(\pi
f_{c})$ of phase qubit is controlled by the external flux $f_{c}$.
%-----------------------------------------------------------
If the phase different $\Theta$ is confined to the period $\pi
f_{c}$, the states of a single qubit are restricted to a two
lowest-energy states $|g\rangle$ and $|e\rangle$.
%-------------------------------------------------------------
%======================================================
%===============================================================
Hence, the Hamiltonian (\ref{a}) can be converted to a spin
representation as follows:
%------------------------------
\begin{eqnarray}\label{b}
\hat{H}_{qu}&=& -\frac{\hbar\omega_{z}}{2}\hat{\sigma}_{z},
\end{eqnarray}
%-----------------------------
where $\omega_{z}\sim\frac{\sqrt{2E_{C}E_{J_{e}}}}{\hbar}$ is the
frequency of the phase qubit, and $\hat{\sigma}_{z}$ is the usual
Pauli spin operator.
%==========================================================
\\ \indent
%--------------------------
The torsional resonator mode is described by a harmonic
oscillator-like Hamiltonian as follows \cite{sss}
%--------------------------------------------------
\begin{eqnarray}
\hat{H}_{mode}
&=&\frac{P^{2}_{\top}}{2I_{m}}+\frac{1}{2}I_{m}\omega^{2}_{v}
\top^{2},
\end{eqnarray}
%----------------------------------------------------------
where $P_{\top}$ is the angular momentum conjugate to $\top$ and
$\top\in[0,2\pi]$.
%------------------------------------------------------
The fluctuations of the angle $\top$ can be described by the
fluctuation $\top_{0}=\sqrt{\hbar/I_{m}\omega_{v}}$ in the ground
state, where $I_{m}$, and $\omega_{v} $ are the rotational moment of
inertia of the torsional resonator, and vibration frequency,
respectively.
%==============================================================
\\ \indent
%---------------
If we put the SC qubit in the torsional resonator, the effective
flux is $f_{c}=f^{'}\sin{\top}$.
%------------------------------------------
Hence, $\top$ is measured relative to the direction of $f_{c}$,
$E_{J}$ is the largest energy scale and coupling between the qubit
and torsional vibrational mode can be a large quantity.
%----------------------------------------------
Thus, we consider the qubit is coupled to resonator mode.
%----------------------------------------------------------
Also, the external magnetic field is parallely applied to the phase
qubit plane.
%---------------------------------------------------------
Then the external field in this case is given by:
$f_{c}=f^{'}(\top-\top_{e})$, where $\top_{e}$ is the angle at
equilibrium measured from the direction of $f_{c}$.
%---------------------------------
In this work, $f^{'}$ is the maximum value of magnetic flux when the
field is perpendicular to the qubit plane.
%------------------------------------------
If $\top_{e}$ is non-zero, the coupling strength is decreased by
factor $\sin \top_{e}$, so consider the angle $\top_{e}=0$.
%----------------------------
However, one concludes that the total Hamiltonian restricted to the
two states is given by:
%-------------------------------------------
\begin{eqnarray}\label{a1}
\hat{H}_{t}&=&
\hbar\omega\hat{a}^{\dag}\hat{a}-\frac{\hbar\omega_{z}}{2}\hat{\sigma}_{z}
+G(\hat{a}+\hat{a}^{\dag})\hat{\sigma}_{x},
\end{eqnarray}
%---------------------
where $(\hat{a}^{\dag})$ $\hat{a}$ and $(\hat{\sigma}^{\dag})$
$\hat{\sigma}$ are the (creation) annihilation operators for the
vibrational mode and phase qubit, respectively, $G=\pi
f^{'}\sqrt{\frac{E_{J}\omega_{z}}{2I_{m} \omega}}$ is the constant
coupling energy between the phase qubit and the torsional resonator,
%---------------
and $\omega=\sqrt{\omega^{2}_{v} +2\pi^{2}f{^{'}}^{2}E_{J}/I_{m}}$
is the torsional oscillator frequency.
%------------------------------------------
Because of the negative sign of the second term, the above
Hamiltonian (\ref{a1}) in the counter-rotating wave approximation
can be written as follows:
%----------------------------------------------------
\begin{eqnarray}\label{a2}
\hat{H}&=&
\hbar\omega\hat{a}^{\dag}\hat{a}-\frac{\hbar\omega_{z}}{2}\hat{\sigma}_{z}
+G(\hat{a}\hat{\sigma}_{-}+\hat{a}^{\dag}\hat{\sigma}_{+}).
\end{eqnarray}
%=================================================================
\\ \indent
%--------------------------
Under the previous  hypothesis, the non-unitary time evolution of
present interaction system with taking into account the dissipative
effect is given through the density operator.
%---------------------------------------------------------
The density operator is described by a master equation of the form:
%-----------------------------------------------------
\begin{eqnarray}\label{a3}
\hbar\hat{\dot{\rho}} &=& i[\hat{\rho},\hat{H}]+
\frac{\hbar\gamma}{2}\Big(\hat{\sigma}_{z}\hat{\rho}\hat{\sigma}_{z}-\hat{\rho}\Big),
\end{eqnarray}
%-------------------------------------------------
where $\gamma$ is a positive parameter that represents a dispersive
reservoir (or phase damping of the qubit system).
%---------------------------------------------------------
It is one of the variables that does not interchange energy with the
qubit system.
%-------------------------------------
Dispersive reservoir of the qubit may occur due to dephasing
interactions that might arise.
%--------------------------------------------------
For instance, the dispersive reservoir may occur from the elastic
collisions in a qubit vapor.
%----------------------------------------------------
The exact solution for Eq.\ref{a3} in the case of a high-$Q$
torsional resonator ($\gamma/G\ll1$) is obtained by analytic method.
%----------------------------------------------------------------
The used analytic method is presented by the dressed-states
representation \cite{c3,c4,c5}.
%--------------------------------------------------------------------
This representation is consisting of the complete set of Hamiltonian
eigenfunctions.
%-------------------------------------------------
Here, the states $|g\rangle$ and $|e\rangle$ are corresponding to
the higher energy and the lower energy, respectively.
%----------------------------------------------------------------
In the invariant subspace spanned by $|e,n+1\rangle$ and
$|g,n\rangle$, the eigenvalues and eigenfunctions of the Hamiltonian
(\ref{a2}) are given by:
%---------------------------------------------
\begin{eqnarray}
\hat{H}|\Psi^{\pm}_{n}\rangle&=&
E^{\pm}_{n}|\Psi^{\pm}_{n}\rangle,\qquad\qquad \hat{H}|e,0\rangle=
-\frac{\hbar\omega_{z}}{2}|e,0\rangle,
\end{eqnarray}
%------------------------------------------------
with the abbreviations
%------------------
$ E_{n}^{{\pm}} =\hbar \omega(n+\frac{1}{2})\pm R_{n}$,
%-----------------
$|\Psi_{n}^{\pm}\rangle =
\beta_{n}^{\mp}|g,n\rangle\pm\beta_{n}^{\pm}|e,n+1\rangle$,
%-----------------------------
$\beta_{n}^{\pm}=\frac{1}{\sqrt{2}}\sqrt{1\mp\frac{\delta}{R_{n}}}$,
$R_{n}=G\sqrt{(n+1)+(\frac{\delta}{G})^{2}}$,
%--------------------
and $\delta=\frac{\hbar(\omega_{z}-\omega)}{2}$ is the detuning
parameter which represents the difference between transition energy
of qubit  and torsional energy.
%==============================================
\\ \indent
%--------------------------------------
To obtain the analytical solution for Eq.\ref{a3}, we suppose that
the qubit is initially in the general superposition of the excited
state and ground state while the torsional resonator is assumed to
be initially in a coherent state, that is
%-------------------------------------------
\begin{eqnarray}
\hat{\rho}(0) &=&
\Big(\cos\theta|e\rangle+e^{i\varphi}\sin\theta|g\rangle\Big)
\Big(\cos\theta|e\rangle+e^{i\varphi}\sin\theta|g\rangle\Big)^{\dagger}
\otimes|\alpha\rangle\langle\alpha|.
\end{eqnarray}
%================================
Consequently, the time-dependent analytical solution for the density
matrix with the initial previous state is given by:
%-------------------------------------------------
\begin{eqnarray}\label{a4}
\hat{\rho}(t) &=&\rho^{00}_{ee}|e,0\rangle\langle
e,0|+\sum^{\infty}_{n,m=0}\Big\{\rho^{n,m}_{gg}(t)|g,n\rangle\langle
g,m|+\rho^{n,m+1}_{ge}(t)|g,n\rangle\langle e,m+1| \nonumber\\&&
+\rho^{n+1,m}_{eg}(t)|e,n+1\rangle\langle
g,m|+\rho^{n+1,m+1}_{ee}(t)|e,n+1\rangle\langle e,m+1|\Big\},
\end{eqnarray}
%=----------------------------------------------------
where all elements of the density matrix (\ref{a4}) are also
functions in $\theta,\varphi,\delta$.
%-----------------------------------------
The elements of the previous density matrix are given by the
following formulas:
%=====================================
%-------------------------------------------------------
\begin{eqnarray*}
%--------------------------------
\hspace{-0.63cm}\rho^{n+1,m+1}_{ee}(t)=\left\{%
\begin{array}{ll}
%-----------------------------------------
x_{n}x^{*}_{m}e^{-iA_{n,m}t}\{2\beta^{+}_{n}\beta^{-}_{n}
\beta^{+}_{m}\beta^{-}_{m}\sin^{2}\theta [\mathbb{E}^{+-}_{n,m}\cos
(R^{+-}_{n,m}t)-\mathbb{K}^{++}_{n,m}\cos (R^{++}_{n,m}t)]\\
%=====================================================
+\xi_{n,m}\cos^{2}\theta[({\beta_{n}^{+}}^{2}{\beta_{m}^{+}}^{2}e^{-iR^{+-}_{n,m}t}
+{\beta_{n}^{-}}^{2}{\beta_{m}^{-}}^{2}e^{iR^{+-}_{n,m}t})\mathbb{E}^{+-}_{n,m}
\\+({\beta_{n}^{+}}^{2}{\beta_{m}^{-}}^{2}e^{-iR^{++}_{n,m}t}
+{\beta_{n}^{-}}^{2}{\beta_{m}^{+}}^{2}e^{iR^{++}_{n,m}t})
\mathbb{K}^{++}_{n,m}]
%==========================================================
\\+\frac{\tilde{\alpha}\beta^{+}_{m}\beta^{-}_{m}\sin2\theta}{2\sqrt{n+1}}
[({\beta_{n}^{+}}^{2}e^{-iR^{+-}_{n,m}t}
-{\beta_{n}^{-}}^{2}e^{iR^{+-}_{n,m}t})
\mathbb{E}^{+-}_{n,m}+({\beta_{n}^{-}}^{2}e^{iR^{++}_{n,m}t}
-{\beta_{n}^{+}}^{2}e^{-iR^{++}_{n,m}t}) \mathbb{K}^{++}_{n,m}]
%===========================================================
\\+\frac{\tilde{\alpha}^{*}\beta^{+}_{n}
\beta^{-}_{n}\sin2\theta}{2\sqrt{m+1}}
[({\beta_{m}^{+}}^{2}e^{-iR^{+-}_{n,m}t}
-{\beta_{m}^{-}}^{2}e^{iR^{+-}_{n,m}t})
\mathbb{E}^{+-}_{n,m}+({\beta_{m}^{-}}^{2}e^{-iR^{++}_{n,m}t}
-{\beta_{m}^{+}}^{2}e^{iR^{++}_{n,m}t})\mathbb{K}^{++}_{n,m}]
%=============================================================
\},&\\
%---------------------
\qquad\qquad\qquad\qquad\quad\quad
%----------------------
\hbox{$\forall$ n$\neq$ m;} \\
%]]]]]]]]]]]]]]]]]]]]]]]]]]]]]]]]]]]]]]]]]]]]]]]]]]]]]]]]]]]]]]]]
%===============================================
|x_{n}|^{2}\{\sin^{2}\theta[\Pi^{+,-}-v^{+,-}]
%===================================
+\xi_{n,n}\cos^{2}\theta [\varpi^{+,-}+v^{+,-}]
%==================================
+\Omega^{+,-}+\Gamma^{+,-}
%====================================================
\},&\\
%---------------------
\qquad\qquad\qquad\qquad\quad\quad
%----------------------
\hbox{$\forall$ n=m.} \\
\end{array}%
\right.
\end{eqnarray*}
%eeeeeeeeeeeeeeeeeeeeeeeeeeeeeeeeeeeeeeeeeeeeeeeeeeeeeeeeeeeeee
\begin{eqnarray*}
\hspace{-0.63cm}\rho^{n,m}_{gg}(t)=\left\{%
\begin{array}{ll}
%--------------------------
x_{n}x^{*}_{m}e^{-iA_{n,m}t}\{
\sin^{2}\theta[({\beta_{n}^{-}}^{2}{\beta_{m}^{-}}^{2}e^{-iR^{+-}_{n,m}t}
+{\beta_{n}^{+}}^{2}{\beta_{m}^{+}}^{2} e^{iR^{+-}_{n,m}t})
\mathbb{E}^{+-}_{n,m}\\+({\beta_{n}^{+}}^{2}{\beta_{m}^{-}}^{2}
e^{iR^{++}_{n,m}t} +{\beta_{n}^{-}}^{2}{\beta_{m}^{+}}^{2}
e^{-iR^{++}_{n,m}t})\mathbb{K}^{++}_{n,m}]\\
%====================================
+2\beta^{+}_{n}\beta^{-}_{n}\beta^{+}_{m} \beta^{-}_{m}\xi_{n,m}
\cos^{2}\theta [\mathbb{E}^{+-}_{n,m}\cos
(R^{+-}_{n,m}t)-\mathbb{K}^{++}_{n,m} \cos (R^{++}_{n,m}t)]
%=====================================
\\+\frac{\tilde{\alpha}\beta^{+}_{n}\beta^{-}_{n}\sin2\theta}{2\sqrt{n+1}}
[({\beta_{m}^{-}}^{2}e^{-iR^{+-}_{n,m}t}
-{\beta_{m}^{+}}^{2}e^{iR^{+-}_{n,m}t})\mathbb{E}^{+-}_{n,m}
+({\beta_{m}^{+}}^{2}e^{-iR^{++}_{n,m}t}
-{\beta_{m}^{-}}^{2}e^{iR^{++}_{n,m}t})\mathbb{K}^{++}_{n,m}]
%==============================================
\\+\frac{\tilde{\alpha}^{*}\beta^{+}_{m}
\beta^{-}_{m}\sin2\theta}{2\sqrt{m+1}}
[({\beta_{n}^{-}}^{2}e^{-iR^{+-}_{n,m}t}
-{\beta_{n}^{+}}^{2}e^{iR^{+-}_{n,m}t})\mathbb{E}^{+-}_{n,m}
+({\beta_{n}^{+}}^{2}e^{iR^{++}_{n,m}t}
-{\beta_{n}^{-}}^{2}e^{-iR^{++}_{n,m}t})
\mathbb{K}^{++}_{n,m}]\},&\\
%================================
\qquad\qquad\qquad\qquad\quad\quad
%----------------------
\hbox{$\forall$ n$\neq$ m;} \\
%===================================================
%-----------------------------------------------------
|x_{n}|^{2}\{\sin^{2}\theta[\varpi^{+,-}+v^{+,-}]
%===================================
+\xi_{n,n}\cos^{2}\theta[\Pi^{+,-}-v^{+,-}]
%==================================
-\Omega^{+,-}-\Gamma^{+,-}\},\\
%================================================
\qquad\qquad\qquad\qquad\quad\quad
%----------------------
\hbox{$\forall$ n=m.} \\
\end{array}%
\right.
\end{eqnarray*}
%eeeeeeeeeeeeeeeeeeeeeeeeeeeeeeeeeeeeeeeeeeeeeeeeeeeeeeeeeeeeeeeee
\begin{eqnarray*}
\hspace{-1.01cm}\rho^{n+1,m}_{eg}(t)=\left\{%
\begin{array}{ll}
%------------------------------------
x_{n}x^{*}_{m}e^{-iA_{n,m}t}\{
\beta^{+}_{n}\beta^{-}_{n}\sin^{2}\theta
[({\beta_{m}^{-}}^{2}e^{-iR^{+-}_{n,m}t}-{\beta_{m}^{+}}^{2}
e^{iR^{+-}_{n,m}t})
\mathbb{E}^{+-}_{n,m}+({\beta_{m}^{+}}^{2}e^{-iR^{++}_{n,m}t}
-{\beta_{m}^{-}}^{2}e^{iR^{++}_{n,m}t}) \mathbb{K}^{++}_{n,m}]\\
%================================================
+\beta^{+}_{m}\beta^{-}_{m}\xi_{n,m} \cos^{2}\theta
[({\beta_{n}^{+}}^{2}e^{-iR^{+-}_{n,m}t}
-{\beta_{n}^{-}}^{2}e^{iR^{+-}_{n,m}t})\mathbb{E}^{+-}_{n,m}
+({\beta_{n}^{-}}^{2}e^{iR^{++}_{n,m}t}
-{\beta_{n}^{+}}^{2}e^{-iR^{++}_{n,m}t})\mathbb{K}^{++}_{n,m}]
%====================================
\\+\frac{\tilde{\alpha}\sin2\theta}{2\sqrt{n+1}}
[({\beta_{n}^{+}}^{2}{\beta_{m}^{-}}^{2}e^{-iR^{+-}_{n,m}t}
+{\beta_{n}^{-}}^{2}{\beta_{m}^{+}}^{2}e^{iR^{+-}_{n,m}t})\mathbb{E}^{+-}_{n,m}
+({\beta_{n}^{+}}^{2}{\beta_{m}^{+}}^{2}e^{-iR^{++}_{n,m}t}
+{\beta_{n}^{-}}^{2}{\beta_{m}^{-}}^{2}e^{iR^{++}_{n,m}t})\mathbb{K}^{++}_{n,m}]
%===================================================
\\+\frac{\tilde{\alpha}^{*}\sin2\theta}{\sqrt{m+1}}
\beta^{+}_{n}\beta^{-}_{n}\beta^{+}_{m}\beta^{-}_{m}[\mathbb{E}^{+-}_{n,m}
\cos(R^{+-}_{n,m}t)-\mathbb{K}^{++}_{n,m}\cos (R^{++}_{n,m}t)]\},&\\
%====================================
\qquad\qquad\qquad\qquad\quad\quad
%----------------------
\hbox{$\forall$ n$\neq$ m;} \\
%******************************************************
%---------------------------------------------
|x_{n}|^{2}\{\beta_{n}^{+}\beta_{n}^{-}
\sin^{2}\theta[({\beta_{n}^{-}}^{2}
-{\beta_{n}^{+}}^{2})e^{\frac{-\gamma
b_{n}t}{2}}\mathbb{E}^{+-}_{n,n}+(
{\beta_{n}^{+}}^{2}e^{-iR^{++}_{n,n}t}-
{\beta_{n}^{-}}^{2}e^{iR^{++}_{n,n}t})\mathbb{K}^{++}_{n,n}]
%=========================================
\\+\beta_{n}^{+}\beta_{n}^{-}
\xi_{n,n}\cos^{2}\theta[({\beta_{n}^{+}}^{2} -{\beta_{n}^{-}}^{2})
e^{\frac{-\gamma b_{n}t}{2}}\mathbb{E}^{+-}_{n,n}+(
{\beta_{n}^{-}}^{2}e^{iR^{++}_{n,n}t}-
{\beta_{n}^{+}}^{2}e^{-iR^{++}_{n,n}t})\mathbb{K}^{++}_{n,n}]
%==============================================================
\\+\frac{{\beta_{n}^{+}}^{2}{\beta_{n}^{-}}^{2}
\sin2\theta(\tilde{\alpha}+\tilde{\alpha}^{*})}{\sqrt{n+1}}
e^{\frac{-\gamma b_{n}t}{2}}\mathbb{E}^{+-}_{n,n}
%=======================================================
\\+\frac{\sin2\theta}{\sqrt{n+1}}[\frac{\tilde{\alpha}}{2}(
{\beta_{n}^{+}}^{4}e^{-iR^{++}_{n,n}t}+
{\beta_{n}^{-}}^{4}e^{iR^{++}_{n,n}t})
-\tilde{\alpha}^{*}{\beta_{n}^{+}}^{2}{\beta_{n}^{-}}^{2}
\cos(R^{++}_{n,n}t)]\mathbb{K}^{++}_{n,n}\},&\\
%==========================================================
\qquad\qquad\qquad\qquad\quad\quad
%----------------------
\hbox{$\forall$ n=m.} \\
\end{array}%
\right.
\end{eqnarray*}
%============================================================
Where
%---------------------------
\begin{eqnarray*}
\begin{array}{ll}
%---------------------------------------------
\rho^{n,m+1}_{ge}= (\rho^{n+1,m}_{eg})^{\dag},\:\:
\rho^{00}_{ee}=e^{-|\alpha|^{2}}\cos^{2}\theta,\:\:
\tilde{\alpha}=\alpha e^{-i\varphi},
\:\:\xi_{n,m}=\frac{|\alpha|^{2}}{\sqrt{(n+1)(m+1)}},\\
%--------------------------------------------
x_{n}=e^{(-\frac{|\alpha|^{2}}{2})}\frac{\alpha^{n}}{\sqrt{n!}},\:\:
R^{+-}_{n,m}=R_{n}-R_{m},\:\:R^{++}_{n,m}=R_{n}+R_{m},\\
%-------------------------------------------------------
\Delta^{+-}_{n,m}=\frac{1}{2}[1
-({\beta^{+}_{n}}^{2}{\beta^{+}_{m}}^{2}
+{\beta^{-}_{n}}^{2}{\beta^{-}_{m}}^{2}
-{\beta^{+}_{n}}^{2}{\beta^{-}_{m}}^{2}
-{\beta^{-}_{n}}^{2}{\beta^{+}_{m}}^{2})],\\
%----------------------------------------------
\Delta^{++}_{n,m}=\frac{1}{2}[1
+({\beta^{+}_{n}}^{2}{\beta^{+}_{m}}^{2}
+{\beta^{-}_{n}}^{2}{\beta^{-}_{m}}^{2}
-{\beta^{+}_{n}}^{2}{\beta^{-}_{m}}^{2}
-{\beta^{-}_{n}}^{2}{\beta^{+}_{m}}^{2})],\\
%------------------------------------------------------
\mathbb{E}^{+-}_{n,m}=e^{-\gamma
t\Delta^{+-}_{n,m}},\:\:\mathbb{K}^{++}_{n,m}=e^{-\gamma
t\Delta^{++}_{n,m}},\:\:
b_{n}=4{\beta_{n}^{+}}^{2}{\beta^{-}_{n}}^{2},\:\:A_{n,m}=\hbar\omega(n-m),\\
%-----------------------------------------------
\Gamma^{+,-}=\frac{\beta_{n}^{+}\beta_{n}^{-}
\sin2\theta}{2\sqrt{n+1}}[{\beta_{n}^{-}}^{2} (\tilde{\alpha}
e^{iR^{++}_{n,n}t}+\tilde{\alpha}^{*}
e^{-iR^{++}_{n,n}t})-{\beta_{n}^{+}}^{2} (\tilde{\alpha}
e^{-iR^{++}_{n,n}t}+\tilde{\alpha}^{*}
e^{iR^{++}_{n,n}t})]\mathbb{K}^{++}_{n,n},\\
%-----------------------------------------------------------------
\Omega^{+,-}=\frac{\beta_{n}^{+}\beta_{n}^{-}\sin2\theta(\tilde{\alpha}
+\tilde{\alpha}^{*})}{2\sqrt{n+1}}
({\beta_{n}^{+}}^{2}-{\beta_{n}^{-}}^{2})e^{\frac{-\gamma
b_{n}t}{2}} \mathbb{E}^{+-}_{n,n},\\
%------------------------------------------------------------------
\Pi^{+,-}=[({\beta_{n}^{+}}^{4}
+{\beta_{n}^{-}}^{4})\sinh(\frac{\gamma
b_{n}t}{2})+2{\beta_{n}^{+}}^{2}{\beta_{n}^{-}}^{2}\cosh(\frac{\gamma
b_{n}t}{2})] \mathbb{E}^{+-}_{n,n},\\
%---------------------------------------------------------------------
\varpi^{+,-}=[({\beta_{n}^{+}}^{4} +{\beta_{n}^{-}}^{4})
\cosh(\frac{\gamma
b_{n}t}{2})+2{\beta_{n}^{+}}^{2}{\beta_{n}^{-}}^{2}\sinh(\frac{\gamma
b_{n}t}{2})]\mathbb{E}^{+-}_{n,n}\:\: \textsf{and}\\
%------------------------------------------------------------
v^{+,-}=2{\beta_{n}^{+}}^{2}{\beta_{n}^{-}}^{2}
\cos(R^{++}_{n,n}t)\mathbb{K}^{++}_{n,n}.
%-------------------------------------------------------------------
\end{array}
\end{eqnarray*}
%==================================================================
\indent
%--------------------
To calculate the asymptotic behavior of the density matrix, we set
$Gt\rightarrow\infty$, as follows:
%--------------------------------------------------------
\begin{eqnarray*}
\hat{\rho}(Gt\rightarrow\infty) &=&\rho^{00}_{ee}|e,0\rangle\langle
e,0|+\frac{1}{2} \sum_{n=0}^{\infty}
|x_{n}|^{2}\Big(\sin^{2}\theta+\xi_{n,n}\cos^{2}\theta\Big)
\nonumber\\&&\times \Big(|e,n+1\rangle\langle
e,n+1|+|g,n\rangle\langle g,n|\Big),
\end{eqnarray*}
%------------------------------------------------------
which is a statistically mixed state.
%wwwwwwwwwwwwwwwwwwwwwwwwwwwwwwwwwwwwwwwwwwwwwwwwwwwwwwwwwwwwwwwwwwwwwwwwwwwwwwwww
\section{The influence of dispersive reservoir on some physical phenomena}
%------------------------------------------------------------
In the current section, we study the existence of some phenomena of
the previous system.
%----------------------------------------------------------------
In an effort to present a numerical characterization, we have
performed some calculations for the properties quantity for a
particular set of parameters, some of which can be considered as
realistic, while some other parameters look perhaps too optimistic.
%-------------------------------------------------------------
%============================================================
\subsection{SC qubit inversion phenomenon}
\begin{figure}
\hspace{-1cm}\includegraphics[height=9cm,width=10cm]{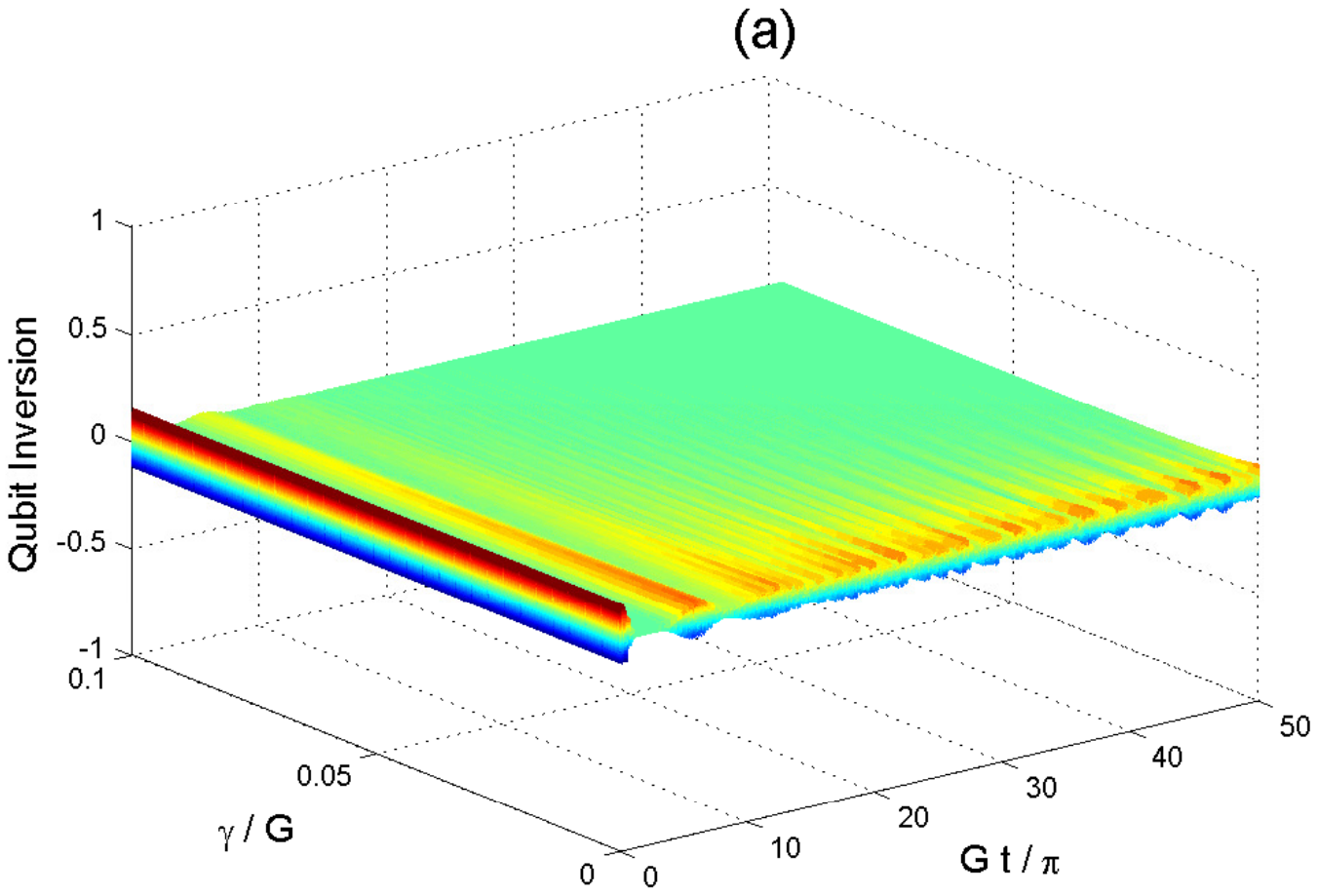}\includegraphics[height=9cm,width=10cm]{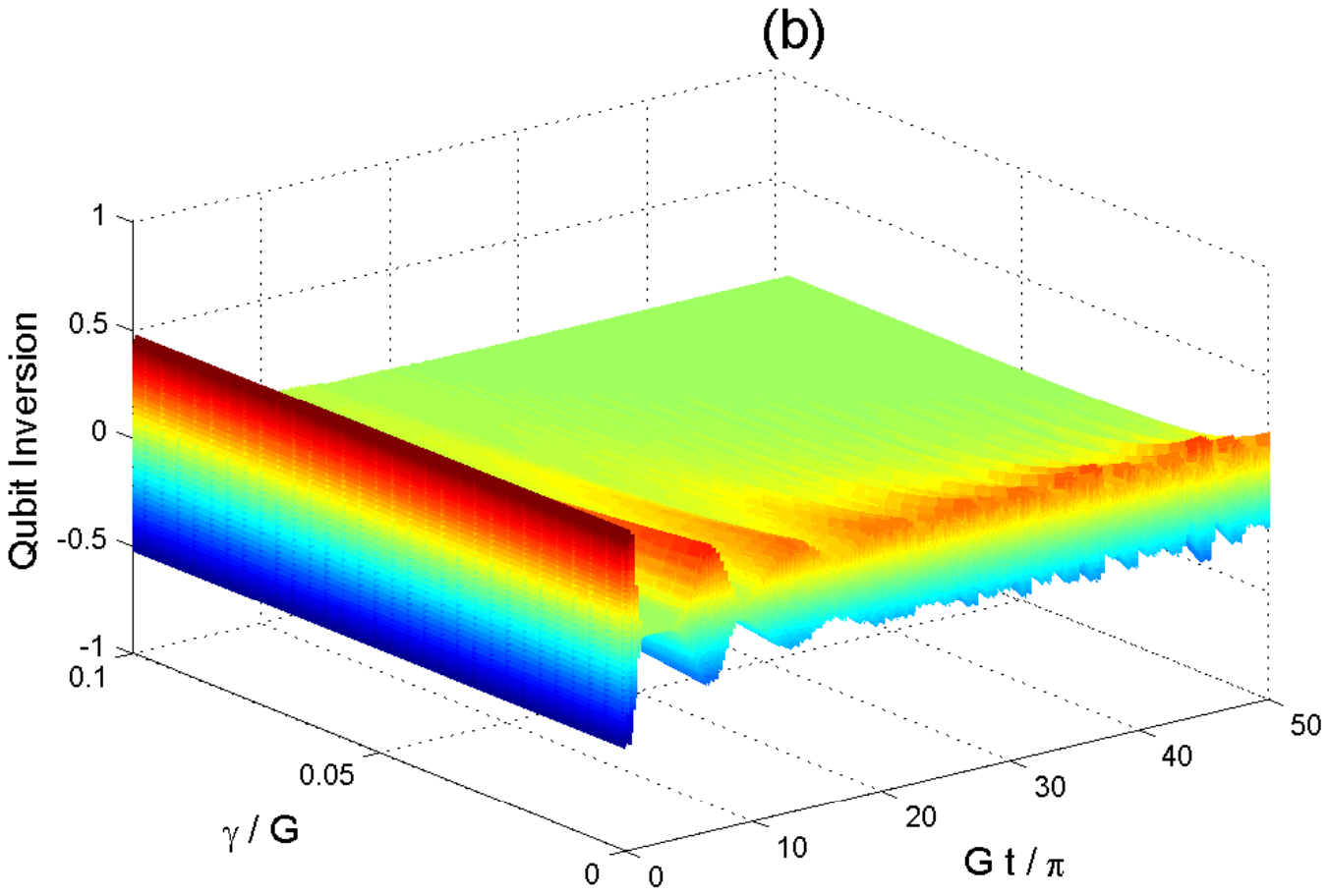}\\
\hspace{-1cm}\includegraphics[height=9cm,width=10cm]{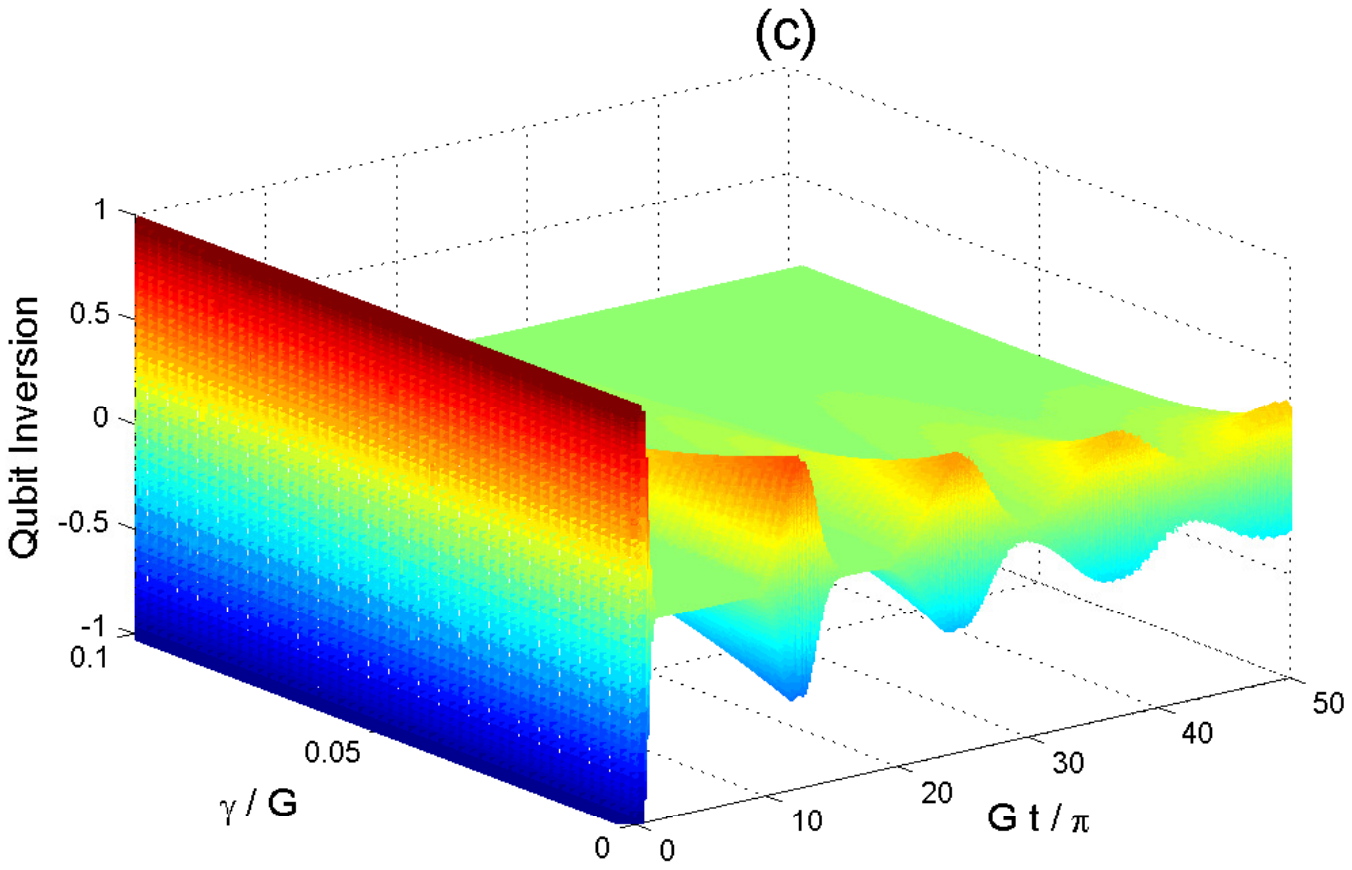}\includegraphics[height=9cm,width=10cm]{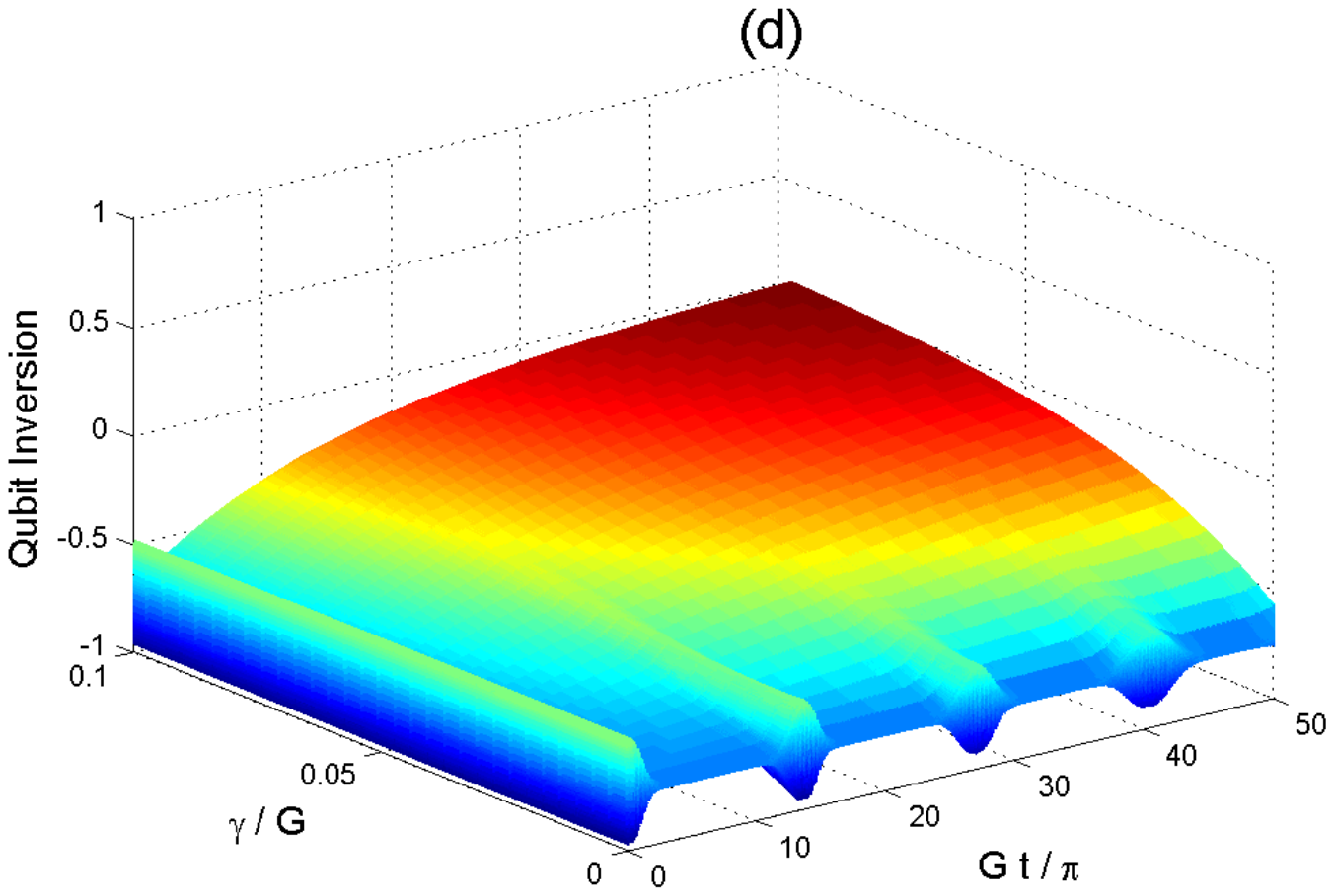}
%----------------------------
\caption{The evolution of SC qubit inversion plotted against the
scaled time $\frac{Gt}{\pi}$ and dispersive reservoir
$\frac{\gamma}{G}$ in the resonance case for $\theta=\frac{\pi}{10}$
in (a), $\theta=\frac{\pi}{3}$ in (b) and $\theta=\frac{\pi}{2}$ in
(c), and off-resonance case for $\theta=\frac{\pi}{3}$ and
$\frac{\delta}{G}=6$ in (d), with the coherence parameter $\alpha=3$
in (a, b, d) and $\alpha=6$ in (c).}\label{w2}
\end{figure}

%----------------
\hspace{0.5cm}
%-------------------------------
The qubit inversion is one of the important dynamic variables of the
properties of quantum mechanics.
%----------------------------------------
It gives information about the behavior of the qubit during the
interaction period.
%-------------------------------------------
Therefore, by using the analytic solution of the master equation and
keeping in mind that the state $|g\rangle$ is the one with higher
energy, we discuss the effect of dispersive reservoir on qubit
inversion, which is given by:
%--------------------------
\begin{eqnarray}
W_{qu}(t)&=& \rho^{00}_{ee}+\sum_{i}\Big(\rho^{ii}_{ee}(t)
-\rho^{ii}_{gg}(t)\Big).
\end{eqnarray}
%==========================================
It is generally accepted that the revivals of inversion appears as a
consequence of quantum coherence which are built up during the
interaction between the resonator and the qubit.
%----------------------------------------------------------
\\ \indent
%---------------------------------
Numerical calculations which indicate the influence of dispersive
reservoir $\frac{\gamma}{G}$ on qubit inversion is exhibited in
Fig.\ref{w2}a, Fig.\ref{w2}b and Fig.\ref{w2}c.
%-----------------------------------------------------
The time evolution of $W_{qu}$ is plotted  as a function of the
scaled time $\frac{Gt}{\pi}$ and $\frac{\gamma}{G}$, in the
resonance case, for a different values of  $\theta$ and $\alpha$.
%-----------------------------------------------------
In the absence of detuning effect and $\frac{\gamma}{G}$, see
Fig.\ref{w2}a and Fig.\ref{w2}b, one finds that $W_{qu}$ fluctuates
around zero, and the collapse and revival phenomena start to appear.
%--------------------------------------------------------
It is seen from Fig.\ref{w2}a and Fig.\ref{w2}b a short period of
collapse after onset of the interaction followed immediately by a
reasonable revival period.
%------------------------------------
After this period of collapse, one can not see another collapse
period because of existence of interference in the fluctuation, so
in these cases of $\theta$ the revival period is not apparent.
%----------------------------------------------------------
The reason for this is due to the value of coherence parameter
$\alpha$.
%--------------------------------------------------------
If $\alpha=6$, the revival and collapse phenomena are more
pronounced than the other case for different values of $\theta$, say
$\frac{\pi}{2}$. This result is observed in Fig.\ref{w2}c.
%------------------------------------------------------------
Furthermore, for many different values of $\theta$, the revival and
collapse phenomena are more pronounced.
%-----------------------------------------------
Also, it is observed from Fig.\ref{w2} that the periodic
oscillations of $W_{qu}$ between $1$ as a maximum value and $-1$ as
a minimum value around $0$.
%==============================================================
\\ \indent
%-------------------------------------------
If the dispersive reservoir parameter is regarded
$\frac{\gamma}{G}\neq0$, the time evolution of $W_{qu}$ is shown in
Figs.\ref{w2}a, Fig.\ref{w2}b and Fig.\ref{w2}c.
%---------------------
At a weak dispersive reservoir, the collapses and revivals
phenomenon is still observed for a short time.
%---------------------
The increase of collapses region slowly due to the influence of the
damping.
%-------------------------------------
But with large values of dispersive reservoir, the amplitude of
revivals is greatly decreased.
%^------------------------------------------
With $\frac{\gamma}{G}>0.1$, all revivals are completely washed out,
and the system is collapsed completely.
%---------------------------------------------
This means that the probability of existence for the qubit in the
state $|g\rangle $ and $|e\rangle$ are equal.
%======================================================
%==============================================================
\\ \indent
%----------------------------
To investigate the influence of detuning parameter
$\frac{\delta}{G}$ and $\frac{\gamma}{G}$ parameter on the behavior
of qubit inversion, we display in Fig.\ref{w2}d the behavior of
inversion in off-resonance case with $\theta=\frac{\pi}{3}$,
$\frac{\delta}{G}=6$ and $\alpha=3$.
%-------------------------------------------
Firstly, for $\frac{\gamma}{G}=0$, one can observe from
Fig.\ref{w2}d, the revivals and collapses phenomena are more
pronounced than the case of $\frac{\delta}{G}=0$, and the curve of
qubit inversion is shifted down.
%-----------------------------------------------------------------
Also, one can note that from  Fig.\ref{w2}d the revival period is
increased, and it can distinguish other periods of revival in that
case, that is in the case of off-resonance for this case of $\theta$
the revival period is more apparent.
%------------------------------------------------------
In fact the revival time is given by: \Big($\tau_{R}=2\pi
G\sqrt{\delta^{2}+|\alpha|^{2}}$\Big) which increases by increasing
$\delta$.
%----------------------------------------------------------
However, we can see from Fig.\ref{w2}d, that the detuning parameter
leads to the state of qubit to tend to $|g\rangle$ more than
$|e\rangle$.
%----------------------------------------------------------
Besides that, if $\frac{\gamma}{G}\neq0$, the collapses and revivals
phenomenon is still observed in a very short time.
%------------------------------------------------------
This phenomenon disappears slowly by increasing time due to the
influence of $\frac{\gamma}{G}$.
%-----------------------------------------------
One can observe rapid deterioration for the revivals of $W_{qu}$ at
the large values of $\frac{\gamma}{G}$.
%--------------------------------------------
From Fig.\ref{w2}d, we observe increasing in the behavior of
$W_{qu}$ curve, after that the value of $W_{qu}$ is almost to zero
at time $Gt\simeq50\pi$.
%-------------------------------------
Moreover, the collapse region has becomes very large.
%------------------------------
However, if we take large values of $\frac{\delta}{G}$ for the
different values of $\theta$, the amplitudes of the fluctuations are
diminished, the collapse periods are elongated, and a large decrease
is happened in the amplitudes of periodic oscillations of maximum
value and minimum value for the inversion.
%--------------------------------------------------------------
%wwwwwwwwwwwwwwwwwwwwwwwwwwwwwwwwwwwwwwwwwwwwwwwwwwwwwwwwwwwwwwwwwwwwwww
\subsection{Dynamical properties of coherence and entanglement phenomena}
%--------------------------------
\hspace{0.5cm}
%-------------------------------
In order to understand the influence of dispersive reservoir of
qubit on some dynamical properties, such as coherence and
entanglement, we use in the following the solution in Eq.\ref{a4}.
%--------------------------------------------------------------
Then we investigate the effect of dispersive reservoir on the
dynamical properties.
%==============================================================
\subsubsection{Dynamical properties of coherence phenomenon}
%------------------------------------------------------
\hspace{0.5cm}
%-------------------------------
In this part, we study the behavior of purity as a measure of the
amount of coherence loss caused by the dispersive reservoir of qubit
and the unitary qubit-vibrational mode interaction.
%----------------------------------------------
The purity of this state is measured by the von Neumann entropy
\cite{b19} as follows:
%-------------------------------------------------------------
\begin{eqnarray}
S_{qu(r)}(t)&=& -\sum_{i=1}^{\infty}\lambda^{i}_{qu(r)}(t)\ln
\lambda^{i}_{qu(r)}(t),
\end{eqnarray}
%-----------------------------------------------------
where $\lambda^{i}_{qu(r)}$ are the instantaneous eigenvalues of the
two subsystems (qubit and torsional resonator).
%-----------------------------------------------
In a bipartite quantum system, the system and subsystem entropies,
at any time $t$, are limited by the following Araki-Lieb
inequality
%-----------------------------------------------
$ |S_{qu}-S_{r}|<S<|S_{qu}+S_{r}|. $
%-----------------------------------------------------
As a result, if the density operator $\rho$ describes a pure state,
the entropy $S$ becomes $0$ and remains constant.
%---------------------------------------------------
This means that, if the system is initially prepared in a pure state
at any time $t>0$, the reduced entropies of the two subsystems are
identical, that is, $S_{qu}(t) = S_{r} (t)$.
%----------------------------------------------------------------------
Otherwise $S>0$, that is, $S_{qu}(t) \neq S_{r} (t)$.
%=============================================================
%-------------------------------------------------------------------
\\ \indent
%-----------------------------------
Here, we begin from pure state, therefore, we either use $S_{r}$ or
$S_{qu}$ in the absence of dispersive reservoir, to measure the
amount of entanglement between the two subsystems, whether in the
resonance or off-resonance cases.
%------------------------------------------
Exactly, $S_{qu}(0)=S_{r}(0)=0$.
%---------------------------------------------------------------------
In this work, we devote our study to the von Neumann entropy for the
reduced density operator of the qubit that can be expressed as
follows:
%-------------------------------------------------------------------
\begin{eqnarray}
S_{qu}(t)&=&-\sum_{i=\pm} \lambda^{i}_{qu}(t)\ln
\lambda^{i}_{qu}(t),
\end{eqnarray}
%---------------------------------------------------------------
where $\lambda^{\pm}_{qu}(t)
=\frac{1}{2}\Big(\rho^{00}_{ee}+\sum_{n=0}^{\infty}
[\rho^{n,n}_{ee}(t)+\rho^{n,n}_{gg}(t)]\Big)
\pm\sqrt{\Lambda^{2}_{n}(t) +\Upsilon^{2}_{n}(t)}$, with $
\Lambda_{n}(t)=\frac{1}{2}\Big(\rho^{00}_{ee}+\sum_{n=0}^{\infty}
[\rho^{n,n}_{ee}(t)-\rho^{n,n}_{gg}(t)]\Big)$\,\,\, and \,\,\,
$\Upsilon_{n}(t)=|\sum_{n=0}^{\infty}\rho^{n,n+1}_{ge}(t)|$.
%================================================================================
\begin{figure}
\hspace{-1cm}\includegraphics[height=9cm,width=10cm]{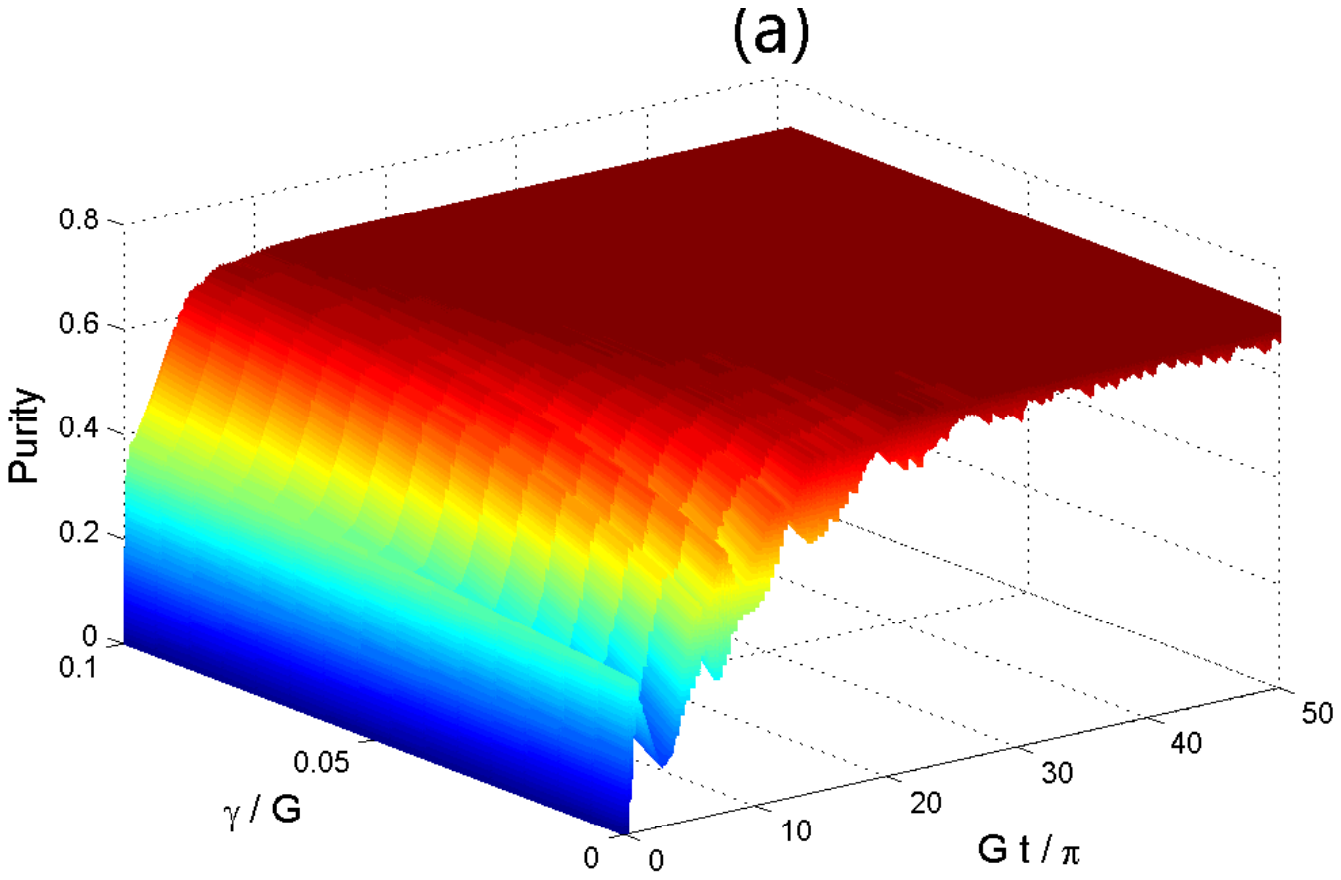}\includegraphics[height=9cm,width=10cm]{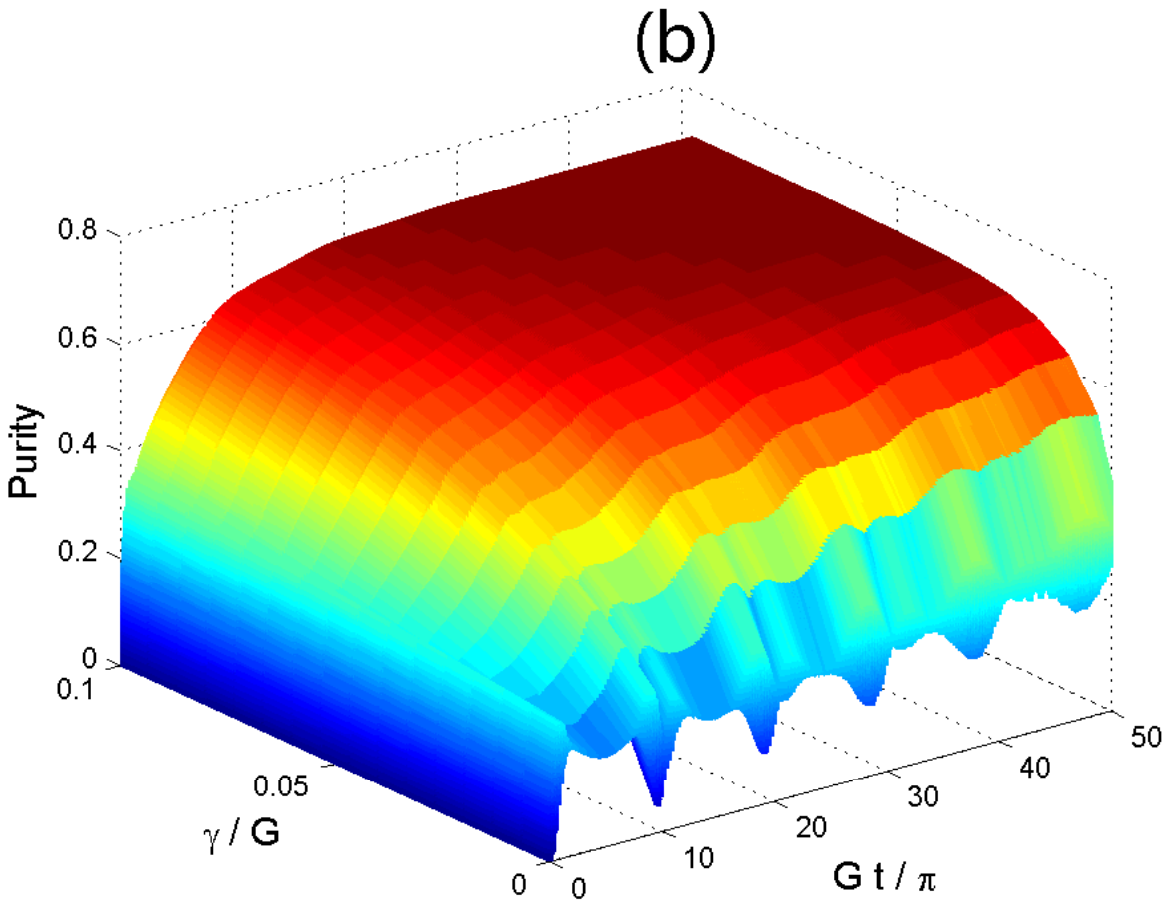}\\
\hspace{-1cm}\includegraphics[height=9cm,width=10cm]{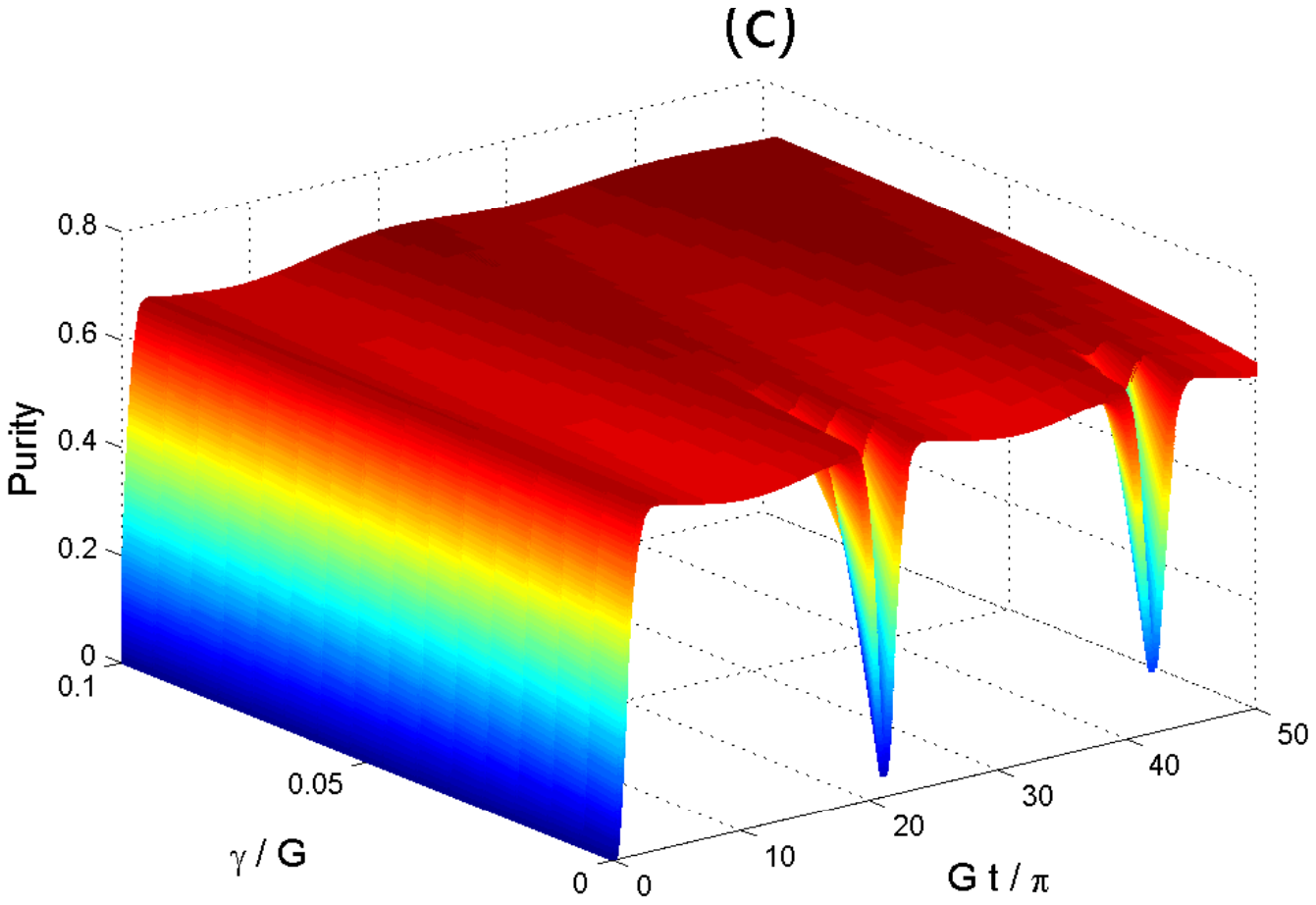}\includegraphics[height=9cm,width=10cm]{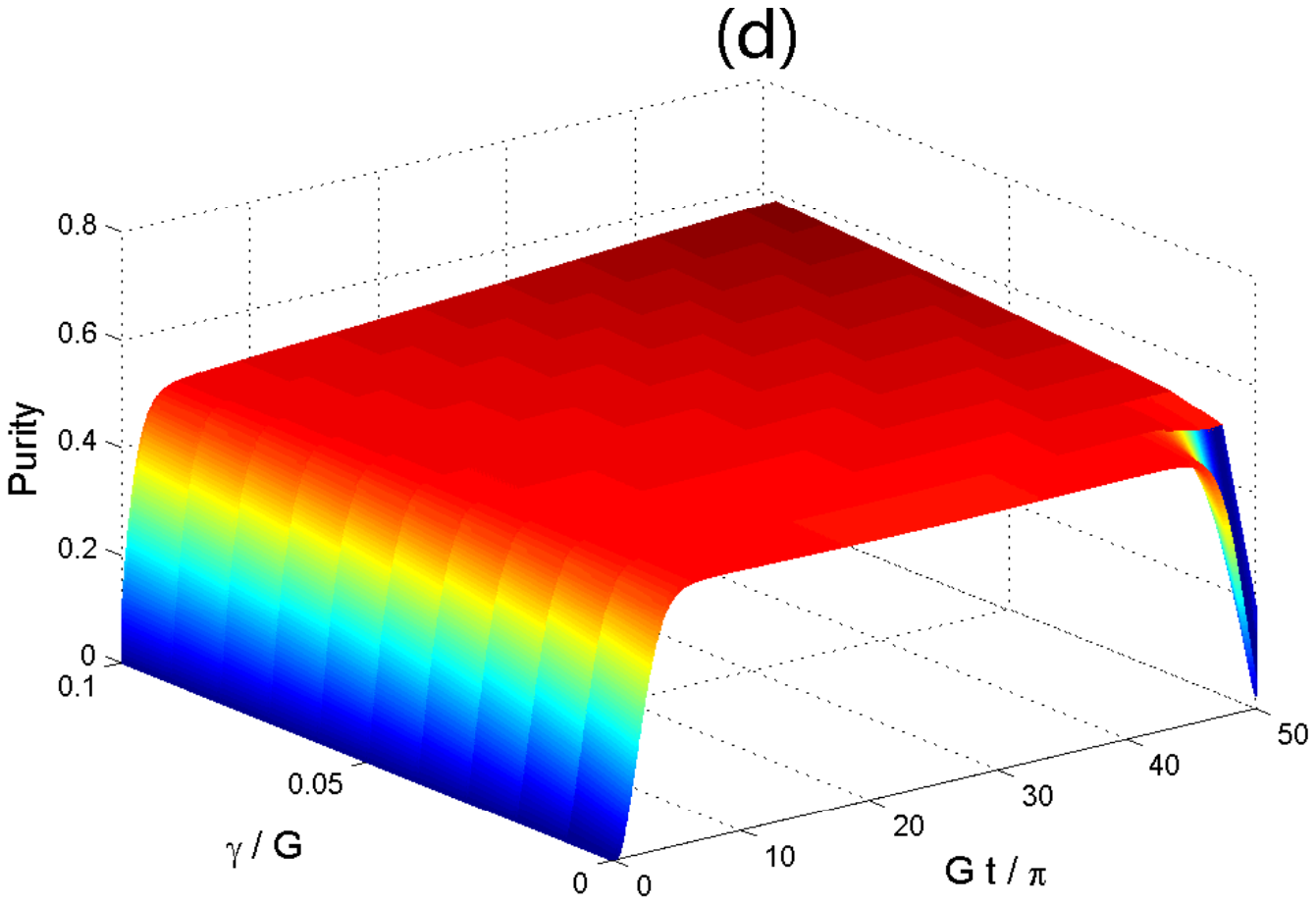}
%----------------------------
\caption{The evolution of purity plotted against the scaled time
$\frac{Gt}{\pi}$ and dispersive reservoir $\frac{\gamma}{G}$ in the
resonance case for $\theta=\frac{\pi}{3}$ in (a), and off-resonance
case for $\theta=\frac{\pi}{3}$ and $\frac{\delta}{G}=4$ in (b),
$\theta=\frac{\pi}{4}$ and $\frac{\delta}{G}=10$ in (c) and
$\theta=\frac{\pi}{8}$ and $\frac{\delta}{G}=25$ in (d), with
$\alpha=3$.}\label{w3}
\end{figure}
%=======================================================
 \indent
%------------------
In our discussion for the influence of dispersive reservoir of qubit
on the purity, the time evolution of the purity against
$\frac{Gt}{\pi}$ and $\frac{\gamma}{G}$, in the resonance case with
$\theta=\frac{\pi}{3}$, $\varphi=0$ and $\alpha=3$, is displayed in
Fig.\ref{w3}a.
%-----------------------------------------------------
From Fig.\ref{w3}a, if $\frac{\gamma}{G}=0$, one can observe that
the behavior of curve of $S_{qu}$ is not periodical.
%-------------------------------------------------
In addition, we see that the time evolution curve of qubit entropy
has local extreme at the revival time
$\Big(t_{R}=\frac{2\pi|\alpha|}{G}=3\sqrt{8I_{m}
\omega/E_{J}\omega_{z}{f^{'}}^{2}}\Big)$, indicating that the qubit
state does not return most closely to its initial state at this
time.
%---------------------------------------------------
Also, one can observe that the value of minima for $S_{qu}$ in
Fig.\ref{w3}a is increased when the time increases.
%------------------------------------------
Finally, the $S_{qu}$ oscillate around its maxima as time becomes
very large.
%===========================================================
%------------------------------------
Besides that, in the presence of the dispersive reservoir
$\frac{\gamma}{G}\neq0$, we cannot use the $S_{qu}$ as a measure for
entanglement but we use it as a measure to the amount of the
coherence loss of the qubit state.
%--------------------------------------------------------------------
At a weak dispersive reservoir, one can find the amplitudes of the
oscillations of entropy diminish rapidly, and the local extreme of
the purity at half-revival time for the mentioned measure are still
apparent, where at this time the qubit and vibration mode are
approximately disentangled.
%--------------------------------------------------------------------
If $\frac{\gamma}{G}=0.1$, the amplitudes of the local extreme of
the purity oscillations are more rapidly deteriorated.
%--------------------------------------------------------------------
Due to damping, the amplitude of the purity of qubit is suppressed.
%----------------------------------
Therefore, the qubit coherence loss may occur due to dephasing
interactions of qubit with the vibration mode.
%---------------------------------------------
However, the purity loses its coherence more than its gain due to
dispersive reservoir, and finally completely loses coherence and
fall into a statistically mixed state.
%-========================= =======================================================
\vspace{0.2cm}\\  \indent
%-----------------------------------------
As well as, we investigate in Fig.\ref{w3}b and Fig.\ref{w3}c,
\ref{w3}d), the evolution of purity in off-resonance case with
$\varphi=0$ and $\alpha=3$ for a different values of $\theta$ and
$\frac{\delta}{G}$.
%--------------------------------------------------------------------
In the absence of $\frac{\gamma}{G}$, see Fig.\ref{w3}b,
Fig.\ref{w3}c and Fig.\ref{w3}d, one can note that the detuning
parameter affects the fluctuations in the temporal evolution for
$S_{qu}$, and results in deterioration of the oscillations.
%--------------------------------------------------------
One can also note from these figures, the local extreme for the
mentioned measure are pronounced and still apparent in all curves.
%--------------------------------------------------------------------
However, we see that the amplitudes of oscillations of purity are
decreasing due to the increase of the value for $\frac{\delta}{G}$
in these different figures, that is the detuning parameter improves
the influence of dispersive reservoir on this phenomenon.
%--------------------------------------------------------------------
In addition, when $\frac{\gamma}{G}\neq0$, one can find from
Fig.\ref{w3}b, Fig.\ref{w3}c and Fig.\ref{w3}d at a weak value of
$\frac{\gamma}{G}$, that the local extreme for the mentioned measure
is about to disappear.
%--------------------------------------------------------------------
Also, one can observe at the high values of $\frac{\gamma}{G}$, that
the amplitudes of oscillations for all curves are deteriorated
completely, and the purity is increased monotonically and no longer
equal zero due to the influence of the dispersive reservoir.
%--------------------------------------------------------------------
\\ \indent
%--------------------
However, from Fig.\ref{w3}c for $\theta=\frac{\pi}{4}$ (phenomenon
of coherent trapping) and $\frac{\delta}{G}=10$, and from
Fig.\ref{w3}d for $\theta=\frac{\pi}{8}$ and $\frac{\delta}{G}=25$.
%-------------------------------------------------
We note that the behavior for two curves of the purity in these
cases of detuning is periodical, and note that the oscillations
periods for two curves are apparent after a large period of time on
the case in which $\frac{\delta}{G}=4$.
%----------------------------------------------------
Then, we conclude that the behavior of curves of purity is also
periodical for the values more than $\frac{\delta}{G}=25$ for the
different values of $\theta$.
%--------------------------------------------------------------------
Finally, when $\frac{\gamma}{G}=1$, we get
$\lambda^{\pm}_{qu}(Gt\rightarrow\infty)=
\frac{1}{2}\pm\frac{1}{2}e^{-|\alpha|^{2}}\cos^{2}{\theta}$, then
$S_{qu}(Gt\rightarrow\infty)\sim\ln2=0.693$, which means that the
asymptotic qubit state is a statistically mixed state, that is the
mixedness of the phase qubit state is complete.
%--------------------------------------------------------------
%wwwwwwwwwwwwwwwwwwwwwwwwwwwwwwwwwwwwwwwwwwwwwwwwwwwwwwwwwwwwwwwwwwwwww
%--------------------------------------
\subsubsection{Dynamical properties of  entanglement phenomenon}
\begin{figure}
\hspace{-1cm}\includegraphics[height=9cm,width=10cm]{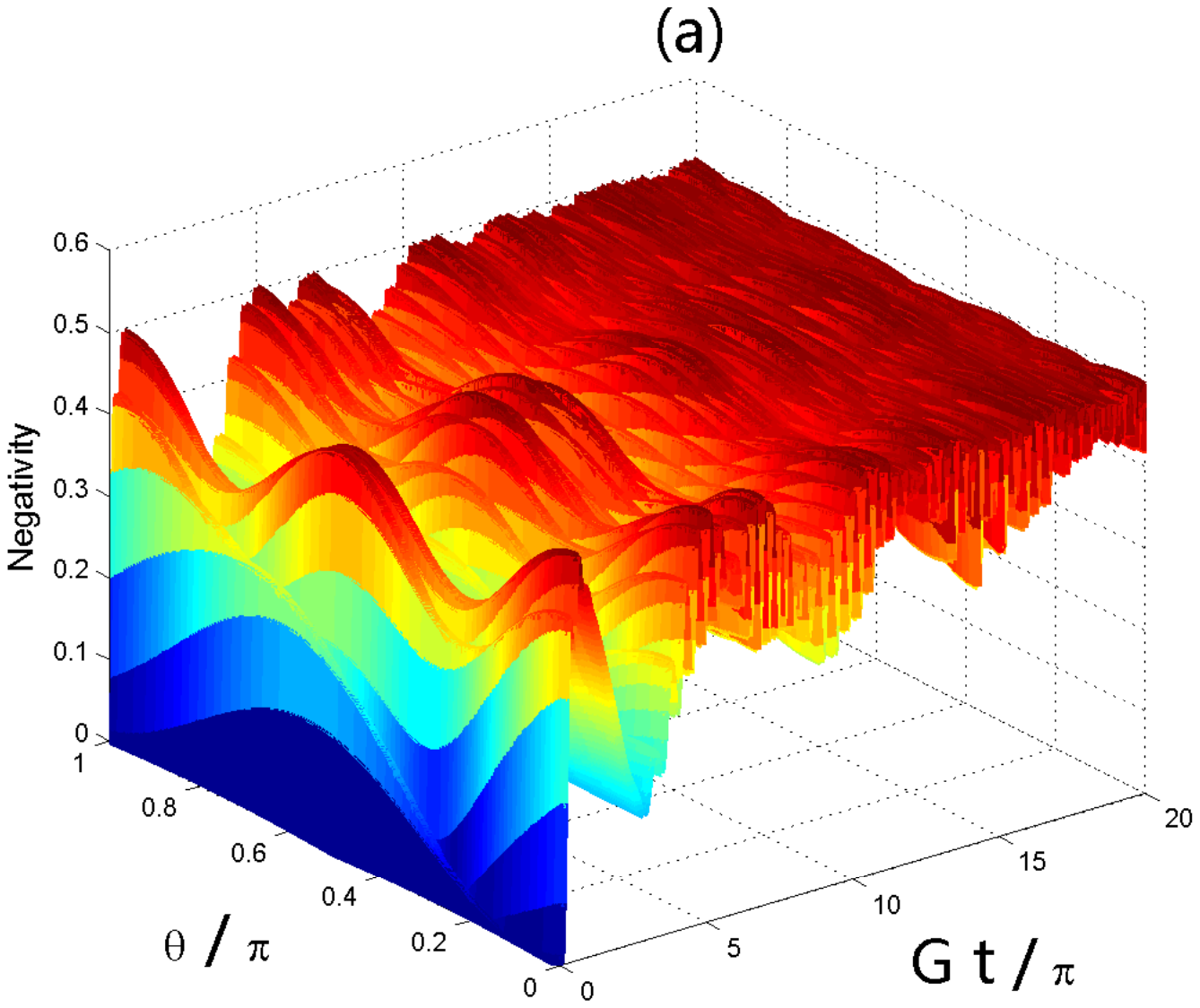}\includegraphics[height=9cm,width=10cm]{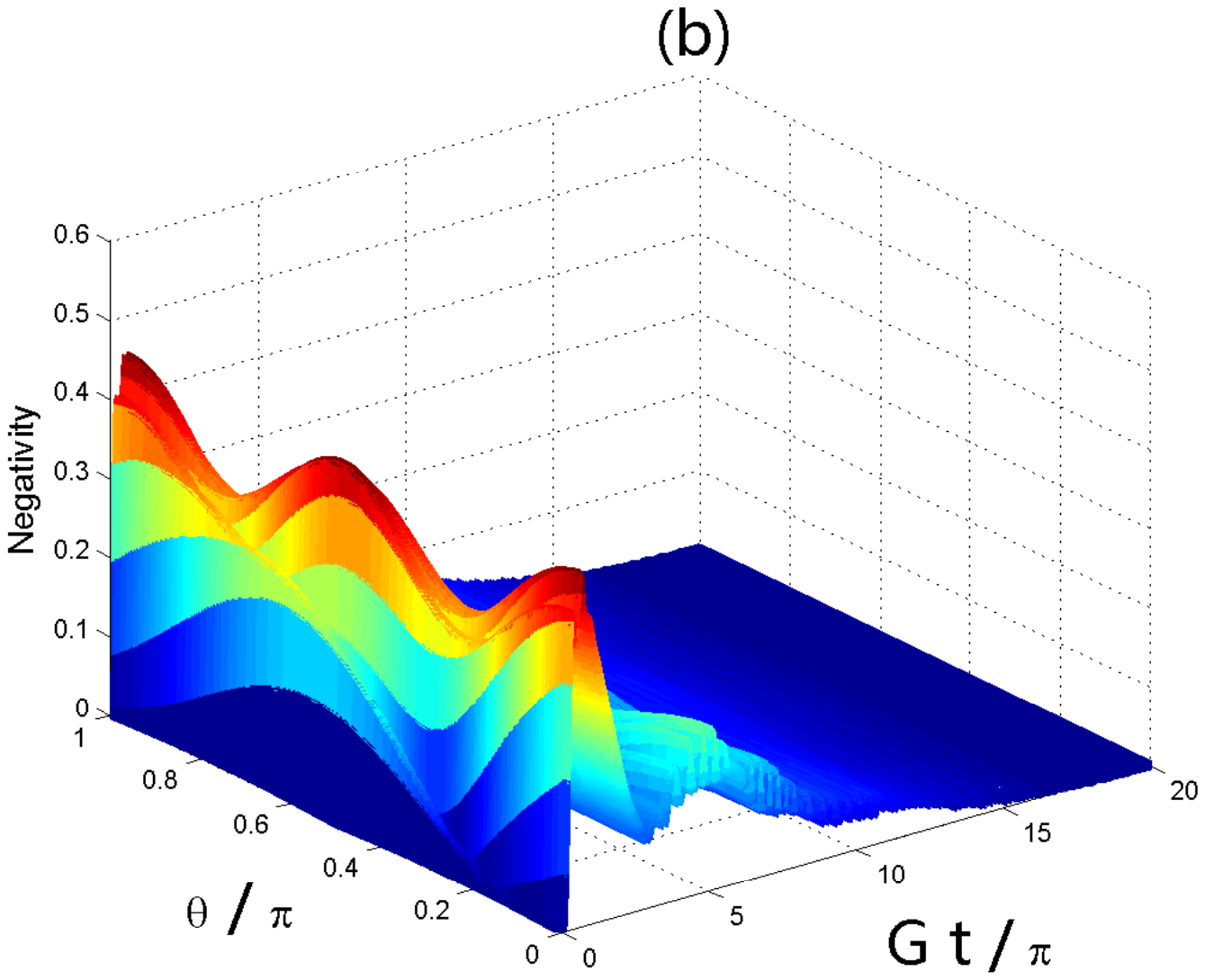}\\
\hspace{-1cm}\includegraphics[height=9cm,width=10cm]{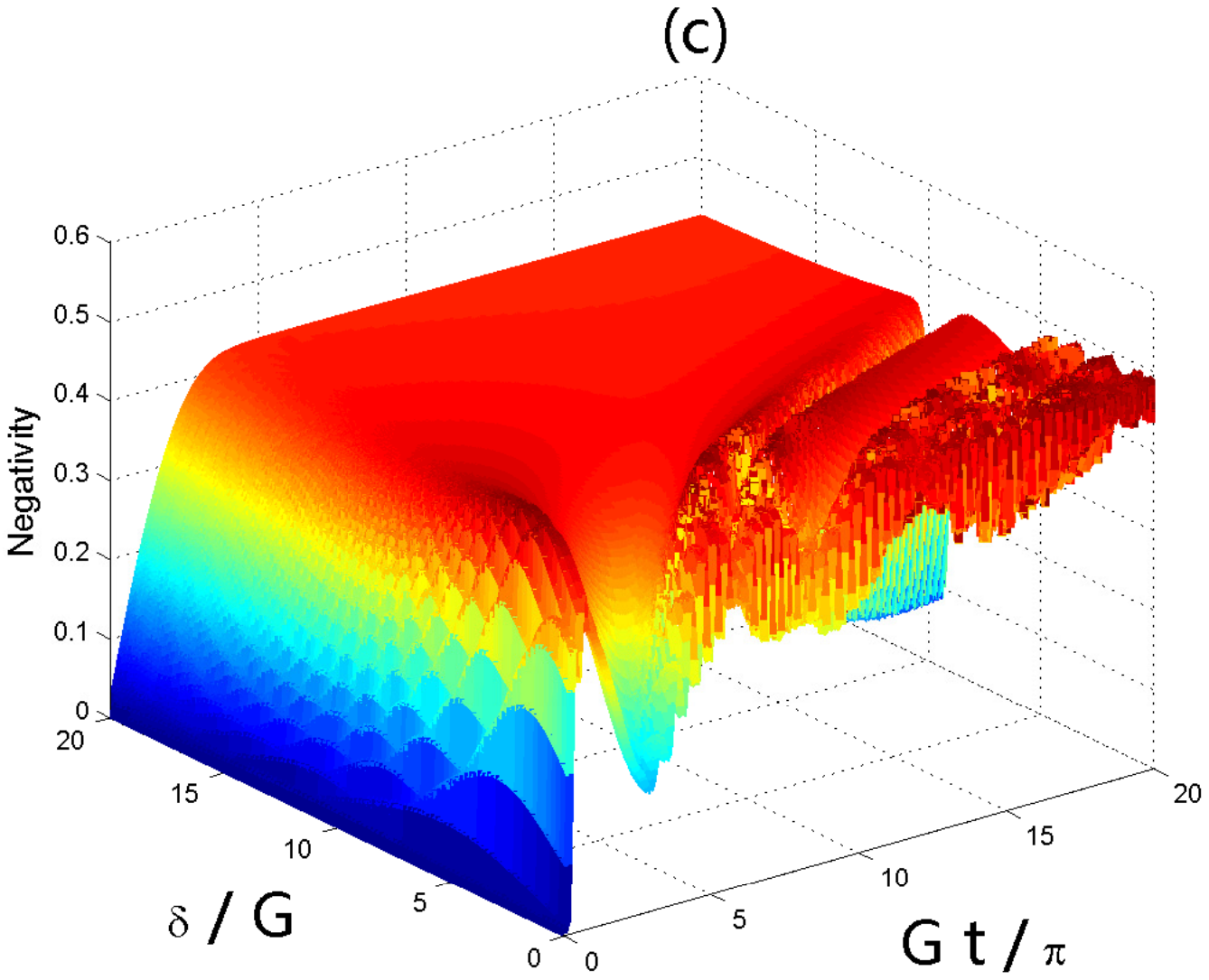}\includegraphics[height=9cm,width=10cm]{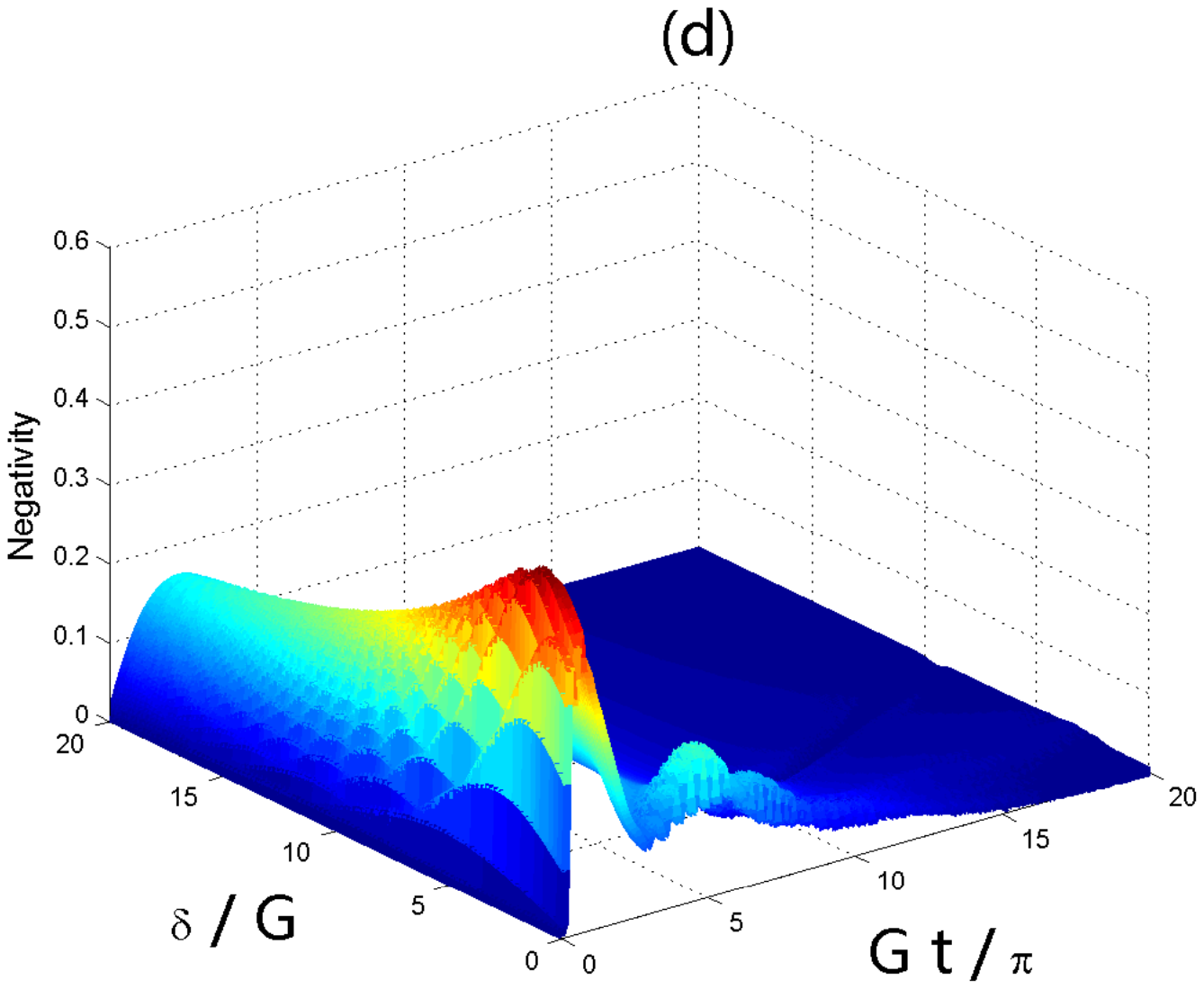}
%----------------------------
\caption{The time evolution of the entanglement $N(\rho)$ in the
resonance case for all $\theta$ and off-resonance case for
$\theta=\frac{\pi}{7}$, with $\varphi=\frac{\pi}{4}$ and $\alpha=3$,
for the different values of dispersive reservoir:
$\frac{\gamma}{G}=0$ in (a, c) and $\frac{\gamma}{G}=0.1$ in (b,
d).}\label{w4}
\end{figure}

%-------------------------------------
\hspace{0.5cm}
%-------------------------------
Because of the existence of dispersive reservoir parameter in
Eq.\ref{a4}, the purity cannot be a measure for the entanglement.
%-------------------------------------
Therefore, we use the negativity $N(\rho)$ to measure the amount of
the entanglement that can be expressed as \cite{b20}
%-------------------------------------------
\begin{eqnarray}
N(\rho) &=& \max\Bigg(0, -\sum_{s}\lambda_{s}\Bigg),
\end{eqnarray}
%------------------------------------------------
where $\lambda_{s}$ is a negative eigenvalue of the partially
transposed density matrix of the vibrational mode-phase qubit.
%--------------------------------------------------
Here, $\lambda_{s}$ is computed by using numerical calculations.
%------------------------------------------------
For an entangled mixed state, $N(\rho)$ is positive.
%------------------------------------------------------
If $N(\rho)=0$, the states are separable.
%------------------------------------------------
Moreover, the negativity is used to estimate the amount of
entanglement of the final state if we start with a pure or mixed
state.
%=========================================================
\\ \indent
%-----------------------------------
The influence of dispersive reservoir of qubit on the negativity as
a function of $\frac{Gt}{\pi}$ and a parameter $\frac{\theta}{\pi}$
in the resonance case when  $\varphi=\frac{\pi}{4}$ and $\alpha=3$
is shown in Fig.\ref{w4}a and Fig.\ref{w4}b.
%------------------------------------------------
From Fig.\ref{w4}a, in the absence of $\frac{\gamma}{G}$, we note
that the local extremes of $N(\rho)$ approximately are regularly
distributed at small times $ Gt\simeq10\pi$ with the different
values of $\theta$.
%------------------------------------------------
After that, the values of minimum for $N(\rho)$ are increased when
the time increases.
%------------------------------------------------
Also, the time evolution curve of $N(\rho)$ has a local extreme at
the revival time $t_{R}$.
%------------------------------------------------
However, we note that, the value of $N(\rho)$ is oscillated around
its maxima for the different values of $\theta$ and $Gt$.
%------------------------------------------------
If we take into account the dispersive reservoir parameter, the
amplitudes of the fast oscillations of the negativity measure
diminish, and $N(\rho)$ arrives to its minima (maxima) value at
half-revival time.
%------------------------------------------------
The negativity has a very strong sensitivity to $\frac{\gamma}{G}$
parameter.
%------------------------------------------------
Therefore, when we take $\frac{\gamma}{G}=0.1$, see Fig.\ref{w4}b,
the $N(\rho)$ quite vanishes at particular times for the different
values of $\theta$.
%------------------------------------------------
In this case, the dispersive reservoir at these particular times,
the entanglement between the vibrational mode and qubit is
completely destroyed.
%-----------------------------------------
If $N(\rho)=0$, the final state shown in Eq.\ref{a4} disentangles
completely and abruptly in just a finite time and evolves to the
vacuum state, and its coherence is lost completely.
%============================================================
%====================================================================
\\ \indent
%------------------------------------------------
To visualize the influence of detuning parameter and
$\frac{\gamma}{G}$ parameter on the temporal evolution for
negativity, the behavior of the negativity in off-resonance case
with $\theta=\frac{\pi}{7}$, $\alpha=3$ and $\varphi=\frac{\pi}{4}$
is displayed in Fig.\ref{w4}c and Fig.\ref{w4}d.
%------------------------------------------------
One can find that, the detuning parameter $\frac{\delta}{G}\neq0$
affects the amplitudes of fluctuations in the evolution of $N(\rho)$
either by increasing or by decreasing on the case in which
$\frac{\delta}{G}=0$ for the different values of $\theta$.
%------------------------------------------------
Therefore, if $\theta=\frac{\pi}{7}$, see Fig.\ref{w4}c, the
amplitudes of oscillations of $N(\rho)$ are less than their
counterparts of the case $\frac{\delta}{G}=0$.
%------------------------------------------------
Because of the detuning parameter, some of oscillations are
deteriorated, the local extremes of the negativity are pronounced
and still apparent in this surface.
%------------------------------------------------
In addition, with $\frac{\gamma}{G}\neq0$, the oscillations of the
local extreme of the $N(\rho)$ are disappeared completely.
%------------------------------------------------
When the dispersive reservoir is increased, say $\frac{\gamma}{G} =
0.1$, see Fig.\ref{w4}d, the amplitude of most oscillations are
suppressed and $N(\rho)$ quite vanishes at particular times.
%------------------------------------------------
This means that, after a particular time, the dispersive reservoir
destroys the entanglement of the global total system state.
%------------------------------------------------
Therefore, one can determine a particular region in which there is
no entanglement between the qubit and the torsional resonator due to
the complete destruction of coherence in the presence of
$\frac{\gamma}{G}$.
%------------------------------------------------
%------------------------------------------------
%wwwwwwwwwwwwwwwwwwwwwwwwwwwwwwwwwwwwwwwwwwwwwwwwwwwwwwwwwwwwwwwwwwwwwww
\section{Conclusion}
%---------------------------
In this article, the system of SC phase qubit coupled to a torsional
resonator is governed by a master equation.
%---------------------------------------
An analytical solution for the master equation is obtained with the
general initial state.
%-----------------------------------
Several physical phenomena have been studied, such as qubit
inversion, purity and negativity in the resonance and off-resonance
cases for many different initial states.
%-------------------------------------
It is found that, the qubit inversion and purity, for the resonance
and off-resonance cases, are quite sensitive to any change
$\frac{\gamma}{G}$ parameter.
%-----------------------------------------------------------
By using the negativity the amount of the entanglement is studied.
%--------------------------------------------
The dispersive reservoir destroys the entanglement of the global
vibrational mode-phase qubit state.
%--------------------------------------------------------------------
In addition, we found that in off-resonance case, the detuning
parameter improve the influence of the damping on qubit inversion,
purity and negativity.
%-------------------------------------
Therefore, one can determine a particular region in which there is
no entanglement between the qubit and the torsional resonator due to
the presence of $\frac{\gamma}{G}$.
%------------------------------------
%wwwwwwwwwwwwwwwwwwwwwwwwwwwwwwwwwwwwwwwwwwwwwwwwwwwwwwwwwwwwwwwwwwwwwwwwwwwwwwwwwwwwwwwwwwww

\end{document}